\documentclass[aps,pre,twocolumn,showpacs]{revtex4}
\usepackage{epsfig}
\usepackage{times}
\draft
\begin{document}

\title{Cooperation enhanced by the difference between
interaction and learning neighborhoods for evolutionary spatial
prisoner's dilemma games}
\author{Zhi-Xi Wu}\thanks{wupiao2004@yahoo.com.cn}
\author{Ying-Hai Wang}\thanks{yhwang@lzu.edu.cn}
\address{Institute of Theoretical Physics, Lanzhou University,
Lanzhou Gansu 730000, China}

\date\today

\begin{abstract}
We study an evolutionary prisoner's dilemma game with two layered
graphs, where the lower layer is the physical infrastructure on
which the interactions are taking place and the upper layer
represents the connections for the strategy adoption (learning)
mechanism. This system is investigated by means of Monte Carlo
simulations and an extended pair-approximation method. We consider
the average density of cooperators in the stationary state for a
fixed interaction graph, while varying the number of edges in the
learning graph. According to the Monte Carlo simulations, the
cooperation is modified substantially in a way resembling a
coherence-resonance-like behavior when the number of learning
edges is increased. This behavior is reproduced by the analytical
results.
\end{abstract}
 \maketitle

\section{introduction}
Cooperation can be found in many places in the realistic world,
from biological systems to economic and social systems
\cite{Dugatkin}. An altruistic action, which benefits others at
the expense of one's own investment, appears to contradict our
understanding of natural selection controlled by selfish
individual behaviors. Thus understanding the conditions for the
emergence and maintenance of cooperative behavior among unrelated
and selfish individuals becomes a central issue in evolutionary
biology \cite{Smith}. In the investigation of this problem the
most popular framework is game theory together with its extensions
involving evolutionary context \cite{Hofbauer,Cressman}. The
prisoner's dilemma (PD), a two-person game in which the players
can choose either cooperation ($C$) or defection ($D$), is a
common paradigm for studying the evolution of cooperation
\cite{Axelrod,Doebeli_1}. In the traditional version of the PD
game, two interacting players are offered a certain payoff, the
reward $R$, for mutual cooperation and a lower payoff, the
punishment $P$, for mutual defection. If one player cooperates
while the other defects, then the cooperator gets the lowest
sucker's payoff $S$, while the defector gains the highest payoff,
the temptation to defect $T$. Thus we obtain $T>R>P>S$. It is easy
to see that defection is the better choice irrespective of the
opponent's selection. For this reason, defection is the only
evolutionarily stable strategy in fully mixed populations of $C$
and $D$ strategies \cite{Hofbauer}.

Since cooperation is abundant and robust in nature, considerable
efforts have been concentrated on exploration of the origin and
persistence of cooperation. During the last decades, five rules,
namely, kin selection \cite{Hamilton}, direct reciprocity
\cite{Axelrod}, indirect reciprocity \cite{NowakNature2005},
network (or spatial) reciprocity
\cite{Ohtsuki2006,Nowak19921,Nowak19922}, and group selection
\cite{Traulsenpnas}, have been found to benefit the evolution of
cooperation in biological and ecological systems as well as within
human societies (for a recent review, see \cite{NowakScience2006}
and references therein). In realistic systems, most interactions
among elements are spatially localized, which makes spatial or
graph models more meaningful. Unlike the other four rules, spatial
games (i.e., network reciprocity) can lead to cooperation in the
absence of any strategic complexity
\cite{Ohtsuki2006,Nowak19921,Nowak19922,Szabo_0} (for a recent
review of evolutionary games on graphs, see \cite{Szabo}). In
spatial evolutionary PD games, the cooperators can survive by
forming large compact clusters, which minimize the exploitation by
defectors. Along the boundary, cooperators can outweigh their
losses against defectors by gains from interactions within the
cluster \cite{Szabo_0,Hauertnature,Hauert}.

In spatial models
\cite{Nowak19921,Nowak19922,Szabo_0,Hauertnature,Hauert}, the
players occupying the vertices of a graph can follow one of the
two pure strategies ($C$ or $D$), and collect payoffs from their
neighbors by playing PD games. Sometimes the players are allowed
to modify their strategies according to an evolutionary rule
dependent on the local payoff distribution. To describe real
systems we can introduce two different graphs
\cite{NowakScience2006}. The ``interaction graph" determines who
plays with whom. The ``replacement graph" (or learning graph)
determines who competes with whom for reproduction, which can be
genetic or cultural. To our knowledge, in most of the existing
works the interaction and replacement graphs are assumed to be
identical. The different roles of these graphs raises a natural
question: How is cooperation affected when the interaction and
replacement graphs are different? Ifti \emph{et al.} \cite{Ifti}
have studied the continuous PD game when the interaction
neighborhood (IN) and learning neighborhood (LN) are different. In
the lattice topology, it was observed that when the neighborhood
sizes for ``interacting" and ``learning" differ by more than
$0.5$, cooperation is not sustainable \cite{Ifti}. Now we wish to
study what happens if the players can follow only one of the two
pure strategies and the LN for the individuals is inhomogeneous.

In this paper, we address these problems by considering an
evolutionary PD game on two layered graphs. The lower layer is the
physical infrastructure on which the interactions are taking place
(interaction layer), and the upper layer represents the
information flows (learning or imitation layer). For the sake of
simplicity, we study the case where the lower interaction layer is
a square lattice. Generally, one can expect that the size of the
LN is larger than that of the IN, which can be understood as
follows. After each round of the game, not only do the interacting
players exchange information about their own payoffs and
strategies, they also share information about their neighbors and
their neighbors' neighbors. To explore the influence of the
difference between the interaction and learning graphs on the
evolution of cooperation, we keep the IN fixed and vary the size
of the LN. In what follows two types of models are systematically
investigated. In the first case (model I), we simply increase the
size of the LN for all the players at the same level. In the
second case (model II), we endow the players with heterogeneous
abilities to obtain information, i.e., some players have a larger
size of LN than others.

\section{model}
We consider the PD game with pure strategies: either $C$ or $D$.
On the interaction layer (a square lattice), each player plays PD
games with its four neighbors and collects a payoff determined by
the strategy-dependent payoff. The total payoff of a certain
player is the sum over all interactions. We assume that a
cooperator pays a cost $c$ for another individual to receive a
benefit $b$ $(b>c)$, and a defector pays no cost and does not
distribute any benefits. Thus the reward for mutual cooperation is
$R=b-c$, the sucker's payoff $S=-c$, the punishment for mutual
defection is $P=0$, and the temptation to defect is $T=b$.
Following \cite{Hauert}, the payoffs are rescaled such that $R=1$,
$T=1+r$, $S=-r$, and $P=0$, where $r=c/(b-c)$ denotes the ratio of
the costs of cooperation to the net benefits of cooperation.

After each round of the game, the players are allowed to inspect
their learning neighbors' payoffs and strategies, and, according
to the comparison, determine which of their strategies to adopt in
the next round. Following previous studies
\cite{Szabo_0,Szabo,Hauert,Szabo_1, Szabo_2,Wu,Szolnoki}, the
evolution of the present system is governed by the adoption of
strategy by a randomly chosen player $i$ and one of its learning
neighbors $j$, namely, the player $i$ will adopt the learning
neighbor's strategy with a probability dependent on the payoff
difference $(U_i-U_j)$ as
\begin{equation}
W = \frac{1}{1 + \exp{[(U_i - U_j)/\kappa]}}\label{rule},
\end{equation}
where $\kappa$ characterizes the noise introduced to permit
irrational choices. $\kappa=0$ and $\kappa\rightarrow \infty$
denote the completely deterministic and completely random
selection of the neighbor's strategy, respectively, while for any
finite positive values $\kappa$ incorporates the uncertainties in
the strategy adoption, i.e., the \emph{better} one's strategy is
readily adopted, but there is a small probability to select the
\emph{worse} one's. The effect of noise $\kappa$ on the stationary
density of cooperators in the spatial PD game has been studied in
detail in Refs. \cite{Szabo_2,Szolnoki}. Since this issue goes
beyond the purpose of the present work, in all our following
studies, we simply fix the value of $\kappa$ to be $\kappa=0.1$.

\begin{figure}
{\epsfxsize=7cm \epsffile{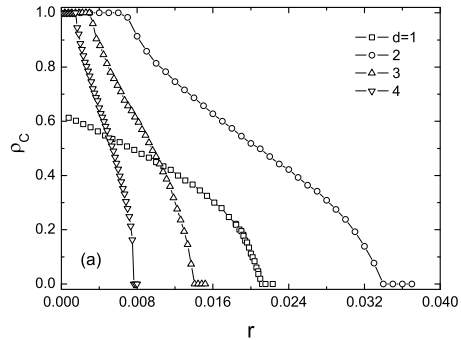} \epsfxsize=7cm
\epsffile{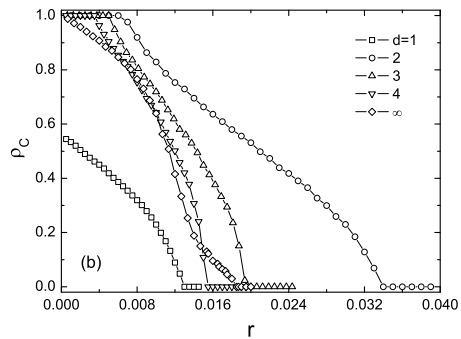} \epsfxsize=7cm \epsffile{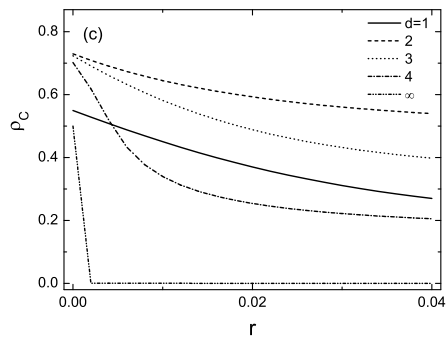}}
\caption{Average density of cooperators, $\rho_C$, as a function
of $r$ for different sizes of the LN on square lattices with
asynchronous (a) and synchronous (b) strategy-updating.
Predictions of $\rho_C$ by the pair approximation are shown in
(c). The cases $d=1, 2, 3, 4$ correspond to, respectively, the
conditions that the learning neighborhoods of the players include
their nearest neighbors, nearest and next-nearest neighbors, and
so on, while $d=\infty$ means that each player can learn from the
whole population. } \label {fig1}
\end{figure}

In both models I and II, the lower interaction graph is a square
lattice with periodic boundary conditions and of size $N=200
\times 200$. For model I, we denote the size of the LN of the
players by $d$, where $d=1,2,\ldots$ indicate, respectively, that
each player can learn (or get payoff and strategy information
after each round) from their nearest neighbors, nearest and
next-nearest neighbors, and so on. For model II,  the upper
learning graph is a scale free network embedded on the underlying
square lattice, which can be constructed according to the
following steps, associated with the lattice embedded scale-free
network (LESFN) model \cite{Rozenfeld}: For each site of the
underlying interaction graph, a prescribed degree $k$ is assigned
taken from a scale-free distribution $P(k)\sim k^{-\gamma}$,
$k\in[4, N)$. A node (say $i$, with degree $k_i$) is picked out
randomly and connected to its closest neighbors until its degree
quota $k_i$ is realized or until all sites up to a distance
$r(k_i)=\min(A\sqrt{k_i}, \sqrt{N})$ have been explored, where $A$
is the territory parameter \cite{Rozenfeld} (in the present work,
we set $A=10$). Duplicate connections are avoided. This process is
repeated for all sites of the underlying lattice \cite{note}.

\section{results}
\begin{figure}
\centerline{\epsfxsize=8cm \epsffile{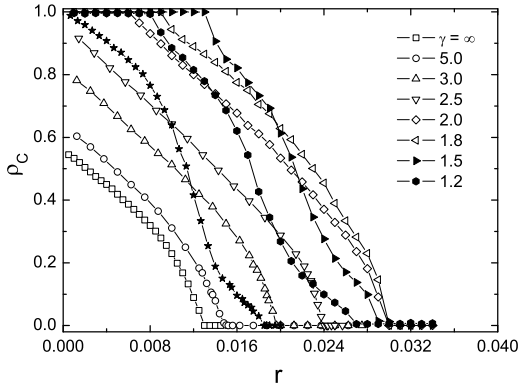}} \caption{Average
density of cooperators, $\rho_C$, as a function of $r$ on square
lattices with synchronous strategy-updating, where the learning
networks are the LESFNs built on the underlying square lattices
with different decay exponents $\gamma$. For the sake of
comparison, the case that each player can learn from the whole
population is also shown by solid stars.} \label {fig2}
\end{figure}

\begin{figure}
\centerline{\epsfxsize=8cm \epsffile{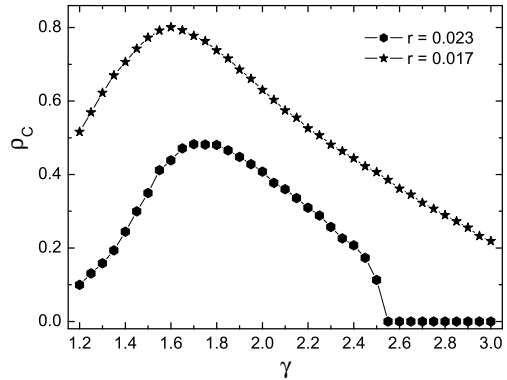}} \caption{Average
density of cooperators, $\rho_C$, as a function of the decay
exponent $\gamma$ of the LESFNs for two special values of
$r=0.017$ and $0.023$.} \label {fig3}
\end{figure}

First, we study the two models by Monte Carlo (MC) simulations
started from a random initial distribution of $C$ and $D$
strategies. By varying the value of $r$, both asynchronous and
synchronous strategy-updating are implemented for model I, and
only synchronous for model II. The total sampling times are
$16000$ MC steps and up to $24000$ for model II when $\gamma<2.0$.
The stationary state is characterized by the average density of
cooperators $\rho_C$ calculated by averaging over the last $4000$
steps when the values of $d$ and $\gamma$ are varied
systematically. All the simulation data shown in Figs. \ref{fig1},
\ref{fig2}, and \ref{fig3} result from an average over either ten
realizations of independent initial strategy configurations (for
model I) or ten realizations of the learning graphs (for model
II).

Let us first discuss the MC results obtained for model I. The
dependence of $\rho_C$ on $r$ in the stationary state for
different sizes of LN, $d$, is illustrated in Figs. \ref{fig1}(a)
and \ref{fig1}(b). For $d=1$, i.e., when the IN and LN are
identical, and with asynchronous strategy updating, we recover the
results of \cite{Hauert}: cooperators persist at substantial
levels if $r$ is sufficiently small [Fig. \ref{fig1}(a)].
Synchronous strategy updating gives rise to a smaller threshold of
$r_c$, beyond which cooperators vanish [Fig. \ref{fig1}(b)]. It is
interesting that for $d=2$, i.e., besides its nearest neighbors a
player can also learn from its next-nearest neighbors, both
asynchronous and synchronous strategy updating lead to
qualitatively as well as quantitatively the same stationary
density of $\rho_C$. For even larger sizes of $d=3$ and $4$,
though the qualitative behaviors are similar, their quantitative
properties are distinct [somewhat greater values of the threshold
$r_c$ in Fig. \ref{fig1}(b) for synchronous dynamics]. In
particular, for $d\rightarrow \infty$, which corresponds to the
case that each player can learn from the whole population,
cooperators cannot persist in the system for any finite positive
values of $r$ when updating asynchronously, whereas they can
maintain at considerable levels if $r$ is very small when updating
synchronously.

In addition to the above points, it is worth pointing out that,
when the LN is larger than the IN, e.g., $d=2,3,$ and $4$, there
arise two absorbing states (all $C$ and all $D$, respectively)
separated by an active state (coexistence of $C$ and $D$) over the
range of $r$, i.e., cooperators can ``wipe out" defectors or
dominate in the system if the players are allowed to get payoff
and strategy information from neighbors further away than just
interacting neighbors only. This is to say, cooperation is
promoted due to the difference between the IN and LN. Figures
\ref{fig1}(a) and \ref{fig1}(b) illustrate clearly the remarkable
enhancement appearing in the case of $d=2$ for both synchronous
and asynchronous dynamics. For synchronous strategy updating,
however, as long as the size of the LN is larger than that of the
IN, the cooperative behavior is always enhanced to some extent as
compared to the case of $d=1$. For asynchronous strategy updating,
however, the promotion of cooperation is only realized in a very
small range of $r$ when $d>2$, and this range decreases with
increasing $d$ and vanishes in the limit of $d\rightarrow \infty$.

The mean-field approximation predicts $\rho_C=0$ for any values of
$0<r<1$ \cite{Szabo,Hauert,Szabo_2}. The nonzero values of
$\rho_c$ (dependent on $r$) cannot be described by the mean-field
approach. To characterize the evolution of $\rho_C$, the more
sophisticated pair approximation provides an analytically
accessible way to determine the corrections from spatial
structural correlation of the players. Instead of the equilibrium
density of $C$ and $D$, the pair approximation considers the
frequency of strategy pairs $C$-$D$ (see the Appendix of Refs.
\cite{Szabo,Hauert} or the Supplementary Information of Ref.
\cite{Hauertnature} for details). However, these existing methods
are prepared for the condition where the interaction and learning
graphs are identical. When these two graphs are different, we
should make some modifications to the original approach.

In the present work, we use an extended pair-approximation (see
the Appendix) to calculate the density $\rho_C$ by varying the
values of $r$ and $d$ for model I. The results are shown in Fig.
\ref{fig1}(c). The extended pair-approximation correctly predicts
the tendencies of the evolution of $\rho_C$, especially for $d=2$,
but significantly underestimates the benefits of spatial
structural effects and the larger size of the LN (than the IN) at
low $r$, whereas it overestimates those at high $r$. Despite this
point, it verifies the above result obtained by MC simulation,
i.e., the most remarkable enhancement of cooperation takes place
at $d=2$. For $d=3$, it fits the synchronous results better than
the asynchronous ones [according to the magnitude relationship
between the curve for $d=3$ and that for $d=1$ in Fig.
\ref{fig1}(a)-Fig. \ref{fig1}(c)], despite the fact that the
pair-approximation is based on the assumption of continuous time,
and hence on asynchronous updating. In particular, for $d=4$, it
correctly predicts the occurrence of an intersection with the
curve for $d=1$ as previously found by MC simulation in Fig.
\ref{fig1}(a). For $d\rightarrow \infty$, it once again correctly
forecasts the extinction of cooperators ($\rho_C=1/2$ for $r=0$,
and $\rho_C=0$ for other any finite positive values of $r$).

\begin{figure}
\centerline{\epsfxsize=8cm \epsffile{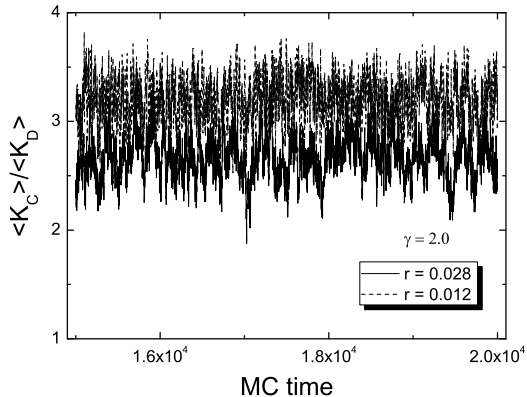}} \caption{Typical
time series of the ratio of average learning degree of cooperators
and defectors, $\langle K_C\rangle / \langle K_D\rangle$, for two
special values of $r=0.012$ and $0.028$ in the stationary state.
The upper LESFN has a decay exponent $\gamma=2.0$.}\label{fig4}
\end{figure}

We now focus our attention on the influence of the heterogeneous
LN on the evolution of cooperation. The MC results obtained for
model II with different values of $\gamma$ are summarized in Fig.
\ref{fig2}. The case $\gamma\rightarrow \infty$ is equivalent to
the case $d=1$ studied in model I. With decreasing $\gamma$
(yielding an increase in the average degree of the LN), the
cooperative level increases gradually until
$\gamma\approx1.7\pm0.2$, where $\rho_C$ reaches its maximum, and
then it gradually decreases as $\gamma$ goes to zero. In finite
size systems for vanishing $\gamma$ in model II, the evolutionary
results  are expected to tend toward the (unattainable) case of
$d\rightarrow \infty$ in model I, since on average the players
have more and more learning neighbors. For the sake of clarity, in
Fig. \ref{fig2} we also show the result for $d\rightarrow \infty$
obtained in model I by solid stars.

Note that, just as was found in model I, if the LN and IN are
different, then cooperation is promoted, and the maximum
enhancement is achieved at a moderate level of the available
information of the LN. Too little information as well as too much
information favors defection. To support this point, we have also
studied the density of cooperators as a function of the size of
the LN of the players (characterized by $\gamma$) for two special
values of $r=0.017$ and $0.023$. The MC results are plotted in
Fig. \ref{fig3}. We can clearly observe that there exactly arise
peak values of $\rho_C$ in the middle range of $\gamma$ for both
$r$, analogously to the so-called coherence resonance
\cite{Traulsen}. In recent research work, many mechanisms are
described that can lead to this coherence resonance phenomenon in
studying the PD game. For example: in Ref. \cite{Traulsen}
additive noise on the classical replicator dynamics can enhance
the average payoff of the population in a resonance-like manner;
By introducing random disorder in the payoff matrix, Perc
\cite{Perc} found a resonance-like behavior of the density of the
cooperators which reaches its maximum at an intermediate disorder.
On static complex networks, Tang \emph{et al.} \cite{Tang}
obtained the result that maximum cooperation level occurs at
intermediate average degree. Ren \emph{et al.} \cite{Ren} studied
the effects of both topological randomness in individual
relationships and dynamical randomness in decision making on the
evolution of cooperation, and found that there exists an optimal
moderate level of randomness, which can induce the highest level
of cooperation. Our result presented here, i.e., enhancing
cooperation by increase of the LN, which resembles a
coherence-resonance-like behavior, provides a different example of
this dynamical phenomenon. It will enrich our knowledge of the
evolution of cooperation in nature.

More recently, Ohtsuki \emph{et al.} \cite{Ohtsuki2007} studied
the evolution of cooperation in the evolutionary spatial PD game,
wherein the interaction graph and replacement (or learning) graph
are separated. They considered three different update rules for
evolutionary dynamics: birth-death, death-birth, and imitation
\cite{Ohtsuki2007}. By both analytical treatment and computer
simulations, they found that under death-birth and imitation
updating, the optimum population structure for cooperators is
given by maximum overlap between the interaction and the
replacement graph, i.e., whenever the two graphs are identical
\cite{Ohtsuki2007}. Any existing difference between these two
graphs will benefit defectors. This result holds for
weak-selection (which means that the payoffs obtained by the
individuals from the game have a slight contribution to their
fitness) and large population size. The ``imitation" updating in
\cite{Ohtsuki2007} is implemented as follows. A random individual
is chosen to update its strategy; it will either stay with its own
strategy or imitate one of the neighbors' strategies proportional
to its fitness. In fact, from this point of view, the update
mechanism (or evolutionary dynamics) of our model, Eq.
(\ref{rule}), can also be regarded as imitation, where the fitness
of each individual is determined by an exponential function of its
payoff obtained from the game, $e^{U/\kappa}$ (Whenever updating
the state of the population, one by one the focal individual and a
randomly chosen neighbor from its LN compete for reproduction
proportional to their fitness according to this function.)
However, we obtain remarkably different result as compared to
\cite{Ohtsuki2007}, i.e., in our model, the difference between the
IN and LN can favor essentially cooperators over defectors
(especially for the case of synchronous updating). Since the
evolutionary outcomes are dependent on the updating rules, and
there are many possible updating dynamics on graphs, we think the
detailed evolutionary rules give rise to this different result in
contrast to that of \cite{Ohtsuki2007}. In addition, in the
present model the fitness of the individuals is closely related to
their payoff, which can be regarded as strong selection, while the
result in \cite{Ohtsuki2007} is obtained in the limit of weak
selection. Thus our present results enrich our knowledge of the
evolution of cooperation in the PD game when the IN and LN are
separated.

Since in model II the players possess an inhomogeneous LN, we
would like to investigate the effect of this heterogeneity on the
players' strategy selection. In Fig. \ref{fig4} we display the
typical stationary-state time series of the ratio of the average
learning degrees of cooperators and defectors (calculated by the
total number of neighbors learning of a certain strategy divided
by the total number of the players adopting this strategy),
$\langle K_C\rangle / \langle K_D\rangle$, for two special values
of $r=0.012$ and $0.028$ (cooperators and defectors dominate in
the two cases, respectively). The upper LESFN has a decay exponent
$\gamma=2.0$. We can observe that in the stationary state the
average learning degree of the cooperators is always larger than
that of the defectors, which indicates that, the more learning
channels the players possess, the greater is the probability they
would cooperate with others.

Finally, we want to point out the difference between our results
and those of Ref. \cite{Ifti}, in which Ifti \emph{et al.} studied
the case where the IN and LN are different in the continuous PD
game, and observed that in the lattice topology, when the
neighborhood sizes for interacting and learning differ by more
than $0.5$, cooperation cannot persist in the population. This is
not the case for the present studied models wherein the players
are pure strategists. Cooperation can be maintained at
considerable levels in the cases where the size of the LN is far
larger than that of the IN [see Figs. \ref{fig1}(a) and
\ref{fig1}(b) for $d=3, 4$ and Fig. \ref{fig2}], and can go so far
as to wipe out defectors for sufficient small $r$ (homogeneous
state $C$ in Figs. \ref{fig1} and \ref{fig2}). In particular, as
long as the strategy updating is implemented synchronously,
cooperation is always promoted essentially when the IN and LN are
different (no matter how large the LN is) as compared when they
are identical.

\section{conclusions}

In summary, we have explored the influence of the difference
between interaction and learning neighborhoods on the evolution of
cooperation. This is done by studying an evolutionary spatial PD
game wherein the interaction and learning graphs of the players
are different. The players are placed on two layered graphs, where
the lower layer is the physical infrastructure on which the
interactions are taking place and the upper layer represents the
skeleton where the payoff and strategy information flow. For the
sake of simplicity, we keep the interaction graph fixed and vary
the size of the neighborhood in the learning graph. Two types of
models have been systematically studied: In model I, we simply
increase the size of interaction neighborhood for all the players
at the same level; and in model II, we endow the players with
heterogeneous ability to obtain information. We performed MC
simulations for both models. For model I, we also use an extended
pair approximation to evaluate the average density of cooperators,
$\rho_C$, and make a comparison with the corresponding results
that follow from our MC simulations.

The main result is that, a difference between the interaction and
learning graphs can promote cooperation substantially. The results
of this mechanism resemble a coherence-resonance-like behavior.
For model I the maximum enhancement is achieved at $d=2$, i.e.,
when the players, in addition to their nearest neighbors, can also
learn from their next-nearest neighbors; for model II, it is
realized at the middle level of the available information of
learning neighbors. Too little learning information favors
defection, but apparently so does too much information (especially
for asynchronous strategy updating). However, as long as the
strategy updating is implemented synchronously, cooperation is
always promoted essentially when choosing a larger size of
neighborhood in the learning graph. This point is also verified by
the extended version of the pair-approximation method. In model
II, where the players possess heterogeneous learning
neighborhoods, we found that the more learning neighbors a player
has, the greater the probability it will cooperate with others.
There are few existing works studying the evolutionary PD game on
networks with distinct interaction and learning neighborhoods.
Thus our present results provide a further perspective on
understanding the emergence and persistence of cooperation in
realistic systems.

In future work, a concise explanation of the mechanism supporting
cooperation should be revealed by more sophisticated analytical
methods. Furthermore, it would be interesting to allow the
interaction neighborhood and/or the learning neighborhood to be
mutable during the process of the dynamics (just as has been done
in the case of the continuous prisoner's dilemma game
\cite{Ifti}), i.e., to study the effects of annealed and quenched
randomness in the interaction and/or learning partnership for
fixed number of coplayers \cite{Szabo_2}. Work along these lines
is in progress.

\section{acknowledgments}
This work was supported by the Fundamental Research Fund for
Physics and Mathematics of Lanzhou University under Grant No.
Lzu05008.

\appendix*
\section{\label{sec:app}extended pair-approximation method}
When the players are pure strategists, an analytical approximation
of the spatial dynamics can be obtained using the pair
approximation (a detailed survey of this technique is given in the
Appendix of the recent review paper \cite{Szabo}, and somewhat
brief yet clear versions can be found in the Supplementary
Information of \cite{Hauertnature} and also in the Appendix of
\cite{Hauert}). Instead of considering the density of strategies
as in well-mixed populations, i.e., in mean-field theory, pair
approximation tracks the densities of strategy pairs. For the
present studied evolutionary PD games, that is to say, we will
first address the probabilities, $p_{c,c}$, $p_{c,d}$, of finding
an individual playing strategy $C$ accompanied by a neighbor
playing $C$ or $D$, respectively. Then the density of $C$ is given
by $\rho_C=p_{c,c}+p_{c,d}$. For more details, we refer the
readers to Refs. \cite{Szabo,Hauert,Hauertnature}. Here we just
make extensions to the approach to study model I, where the
interaction and learning graphs are different. As an example, we
will consider the case of $d=4$ (extensions to other cases are
straightforward).

\begin{figure*}
\centerline{\epsfxsize=18cm \epsffile{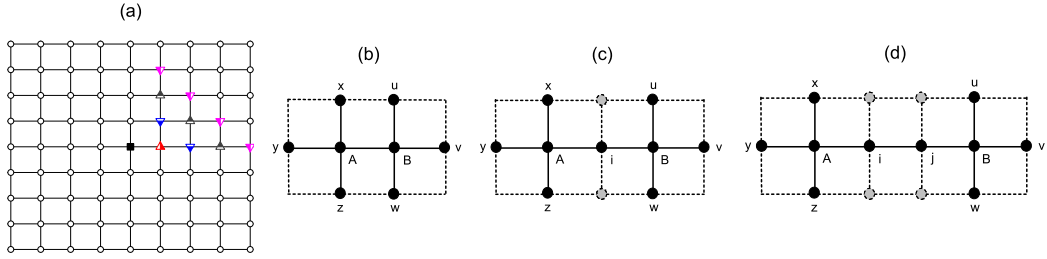}} \caption{(Color
online) Illustration of the lower interaction graph (a square
lattice) and the central site (fully-filled square) has a learning
neighborhood of size $d=4$ [only shown those neighbors falling in
the first quadrant] (a), and the corresponding schemes used for
the pair approximation with involved sites $A$, $x$, $y$, $z$,
$i$, $j$, $u$, $v$, $w$, and $B$ [(b)-(d)]. These schemes are used
to determine changes in the pair configuration probabilities
$p_{A,B}\rightarrow p_{B,B}$ (b), $p_{A,i}\rightarrow p_{B,i}$ (c)
and (d).} \label {fig5}
\end{figure*}

For $d=4$, each player interacts with its four nearest neighbors
on a square lattice, but can learn from those neighbors with
longer (Euclidean) distance up to $4$. Since these learning
neighbors satisfy the condition of rotation symmetry, we will
consider only those neighbors falling in the first quadrant [see
Fig. \ref{fig5}]. Whenever a randomly chosen site $A$ updates its
strategy, a neighbor $B$ is randomly selected from its learning
neighborhood as a reference. Their common neighbors (if any) as
well as their respective neighbors are considered to be
independent by the pair approximation. Thus, when the selected
reference is its nearest (next-nearest) neighbor, we will refer to
the configuration Fig. \ref{fig5}(b) (Fig. \ref{fig5}(c)); and for
other cases we will refer to the configuration Fig. \ref{fig5}(d).
Assuming the selected learning neighbor $B$ is $A$'s third
neighbor (next-next-nearest neighbor), then we will use the scheme
Fig. \ref{fig5}(d) to calculate changes in the pair configuration
probabilities $p_{A,i}\rightarrow p_{B,i}$.

The payoffs $P_A$ and $P_B$ of $A$ and $B$ are determined by
accumulating the payoffs in interactions with their neighbors
$x,y,z,i$ and $u,v,w,j$, respectively. The pair approximation is
completed by determining the evolution of the pair configuration
probabilities, i.e., the probability that the pair $p_{A,i}$
becomes $p_{B,i}$ :
\begin{eqnarray}
p_{A,i\rightarrow B,i}=\sum_{xyz}\sum_{uvw}\sum_{ij}f(P_B-P_A)\times \nonumber \\
\frac{p_{x,A}p_{y,A}p_{z,A}p_{A,i}p_{i,j}p_{j,B}p_{u,B}p_{v,B}
p_{w,B}}{p_A^3p_B^3p_ip_j}\label{configuration},
\end{eqnarray}
where the transition probability $f(P_B-P_A)$[see Eq.
(\ref{rule})] is multiplied by the configuration probability and
summed over all possible configurations. If $B$ succeeds in taking
over site $A$, the following pair configuration probabilities
increase: $p_{x,B},p_{y,B},p_{z,B},p_{B,i}$, while the
probabilities $p_{x,A},p_{y,A},p_{z,A},p_{A,i}$ decrease. It is
easy to analyze the other cases of $B$ (i.e., not the third
neighbor of $A$), which lead to only a slightly different form of
Eq. (\ref{configuration}). All these changes result in a set of
ordinary differential equations:

\begin{widetext}
\begin{eqnarray}
\dot{p}_{c,c}&=&\sum_{xyz}
\{[n_c(x,y,z)+1]h_1+[(n_c(x,y,z)+1)\frac{p_{c,c}}{p_c}+
n_c(x,y,z)\frac{p_{d,d}}{p_d}]h_2+[(n_c(x,y,z)+1)(\frac{p_{c,c}^2}{p_c^2}
+\frac{p_{c,d}^2}{p_cp_d})+\nonumber \\
&&n_c(x,y,z)(\frac{p_{d,d}^2}{p_d^2}+\frac{p_{d,d}p_{c,c}}{p_dp_c})]h_3\}
p_{d,x}p_{d,y}p_{d,z}\times
\sum_{u,v,w}p_{c,u}p_{c,v}p_{c,w}f(P_c(u,v,w)-P_d(x,y,z))-\nonumber \\
&&\sum_{xyz}\{n_c(x,y,z)h_1+[(n_c(x,y,z)+1)\frac{p_{c,c}}{p_c}
+n_c(x,y,z)\frac{p_{d,d}}{p_d}]h_2+[(n_c(x,y,z)+1)(\frac{p_{c,c}^2}{p_c^2}
+\frac{p_{c,c}p_{d,d}}{p_cp_d})+\nonumber \\
&&n_c(x,y,z)(\frac{p_{d,d}^2}{p_d^2}+\frac{p_{c,d}^2}{p_dp_c})
]h_3\}p_{c,x}p_{c,y}p_{c,z}\times
\sum_{u,v,w}p_{d,u}p_{d,v}p_{d,w}f(P_d(u,v,w)-P_c(x,y,z))\\
\dot{p}_{c,d}&=&\sum_{xyz}\{[(1-n_c(x,y,z))]h_1+[(1-n_c(x,y,z))\frac{p_{c,c}}{p_c}
+(2-n_c(x,y,z))\frac{p_{d,d}}{p_d}]h_2+[(1-n_c(x,y,z))(\frac{p_{c,c}^2}{p_c^2}
+\frac{p_{c,d}^2}{p_cp_d})+\nonumber\\
&&(2-n_c(x,y,z))(\frac{p_{d,d}^2}{p_d^2}+\frac{p_{d,d}p_{c,c}}{p_dp_c})]h_3\}
p_{d,x}p_{d,y}p_{d,z}\times\sum_{u,v,w}p_{c,u}p_{c,v}p_{c,w}
f(P_c(u,v,w)-P_d(x,y,z))-\nonumber \\
&&\sum_{xyz}\{[(2-n_c(x,y,z))]h_1+[(1-n_c(x,y,z))\frac{p_{c,c}}{p_c}+(2-n_c(x,y,z))
\frac{p_{d,d}}{p_d}]h_2+[(1-n_c(x,y,z))(\frac{p_{c,c}^2}{p_c^2}
+\frac{p_{c,c}p_{d,d}}{p_cp_d})+\nonumber\\
&&(2-n_c(x,y,z))(\frac{p_{d,d}^2}{p_d^2}+
\frac{p_{c,d}^2}{p_dp_c})]h_3\}p_{c,x}p_{c,y}p_{c,z}\times
\sum_{u,v,w}p_{d,u}p_{d,v}p_{d,w}f(P_d(u,v,w)-P_c(x,y,z)),
\end{eqnarray}
\end{widetext}
where $n_c(x,y,z)$ is the number of cooperators among the
neighbors $x,y,z$, and $P_c(x,y,z)$ and $P_d(x,y,z)$ specify the
payoffs of a cooperator (defector) interacting with the neighbors
$x,y,z$ plus a defector (cooperator). $h_1$, $h_2$, and $h_3$
denote the probabilities of selecting the first, second, and
$\geq$ third next neighbors as references, respectively (see Table
\ref{tab:table1}). For simplicity, the above two equations omit
the common factor $2p_{c,d}/(p_c^3p_d^3)$ which is inessential
\cite{Hauert}. In combination with the symmetry condition
$p_{c,d}=p_{d,c}$ and the constraint
$p_{c,c}+p_{c,d}+p{d,c}+p_{d,d}=1$, the above equations can be
treated either by numerical integration or by setting $\dot
p_{c,c}=\dot p_{c,d}=0$ and solving for $p_{c,c}$ and $p_{c,d}$.
Then the equilibrium density of cooperators is obtained from
$p_c=p_{c,c}+p_{c,d}$.

\begin{table}
\caption{\label{tab:table1}The probabilities of selecting the
nearest neighbors as references, $h_1$, the next-nearest
neighbors, $h_2$, and the remaining cases, $h_3$, for different
sizes of the learning neighborhood.}
\begin{ruledtabular}
\begin{tabular}{ccccc}
 &$h_1$&$h_2$&$h_3$&\\
\hline
$d=1$ & 1 & 0 & 0\\
$d=2$ & 1/3 & 2/3 & 0\\
$d=3$ & 1/6 & 1/3 & 1/2\\
$d=4$\footnote{As an example, the values of $h_1$, $h_2$, $h_3$,
for $d=4$ can be easily counted out by using the symbolized sites
in Fig. \ref{fig5}(a).} & 1/10 & 1/5 & 7/10\\
$d\rightarrow\infty$ & $\approx 0$ & $\approx 0$& $\approx 1$\\
\end{tabular}
\end{ruledtabular}
\end{table}

\end{document}